\documentclass[aps,pra,reprint,superscriptaddress,amsmath,amssymb]{revtex4-1}

\usepackage{CJK}
\usepackage{braket}
\usepackage{tikz}
\usepackage[colorlinks]{hyperref}
\renewcommand{\Im}{\mathrm{Im}}
\renewcommand{\Re}{\mathrm{Re}}
\allowdisplaybreaks

\begin{document}

\begin{CJK*}{UTF8}{gbsn}

\title{Quantum circuit for measuring an operator's generalized expectation values and its applications to non-Hermitian winding numbers}

\author{Ze-Hao Huang (黄泽豪)}
%\thanks{}
\affiliation{National Laboratory of Solid State Microstructures and School of Physics, Nanjing University, Nanjing 210093, China}
\affiliation{Collaborative Innovation Center of Advanced Microstructures, Nanjing University, Nanjing 210093, China}


\author{Peng He (何鹏)}
%\thanks{}
\affiliation{National Laboratory of Solid State Microstructures and School of Physics, Nanjing University, Nanjing 210093, China}
\affiliation{Collaborative Innovation Center of Advanced Microstructures, Nanjing University, Nanjing 210093, China}

\author{Li-Jun Lang (郎利君)}
\email{ljlang@scnu.edu.cn}
%\thanks{}
\affiliation{Guangdong Provincial Key Laboratory of Quantum Engineering and Quantum Materials, School of Physics and Telecommunication Engineering, South China Normal University, Guangzhou 510006, China}

\author{Shi-Liang Zhu (朱诗亮)}
\email{slzhu@nju.edu.cn}
%\thanks{}
\affiliation{Guangdong Provincial Key Laboratory of Quantum Engineering and Quantum Materials, School of Physics and Telecommunication Engineering, South China Normal University, Guangzhou 510006, China}
\affiliation{Guangdong-Hong Kong Joint Laboratory of Quantum Matter, Frontier Research Institute for Physics, South China Normal University, Guangzhou 510006, China}

\date{\today}

\begin{abstract}
We propose a general quantum circuit based on the {\textsc{swap}} test for measuring the quantity $\langle \psi_1 | A | \psi_2 \rangle$ of an arbitrary operator $A$ with respect to two quantum states $|\psi_{1,2}\rangle$.
This quantity is frequently encountered in many fields of physics, and we dub it the generalized expectation as a two-state generalization of the conventional expectation.
We apply the circuit, in the field of non-Hermitian physics, to the measurement of generalized expectations with respect to left and right eigenstates of a given non-Hermitian Hamiltonian.
To efficiently prepare the left and right eigenstates as the input to the general circuit, we also develop a quantum circuit via effectively rotating the Hamiltonian pair $(H,-H^\dagger)$ in the complex plane.
As applications, we demonstrate the validity of these circuits in the prototypical Su-Schrieffer-Heeger model with nonreciprocal hopping by measuring the Bloch and non-Bloch spin textures and the corresponding winding numbers under periodic and open boundary conditions (PBCs and OBCs), respectively.
The numerical simulation shows that non-Hermitian spin textures building up these winding numbers can be well captured with high fidelity, and the distinct topological phase transitions between PBCs and OBCs are clearly characterized.
We may expect that other non-Hermitian topological invariants composed of non-Hermitian spin textures, such as non-Hermitian Chern numbers, and even significant generalized expectations in other branches of physics would also be measured by our general circuit, providing a different perspective to study novel properties in non-Hermitian as well as other physics realized in qubit systems.

\end{abstract}

%\keywords{}

\maketitle

\end{CJK*}

\section{Introduction} \label{sec: introduction}

Since the theoretical prediction and experimental observation of unique features in non-Hermitian systems \cite{Ashida2020,BergholtzKunst2021},
such as the parity-time-reversal symmetry breaking \cite{Bender1998,AGuo2009,BPeng2014,BGZhu2014,JMLi2019,ZJRen2022},
the breakdown of conventional bulk-boundary correspondence \cite{Lee2016,Leykam2017,LXiao2017,YXiong2018,MartinezAlvarez2018a,Kunst2018,HTShen2018,SYYao2018a,SYYao2018b,ZPGong2018,CHYin2018,Yokomizo2019,LJin2019,CHLee2019a,KZhang2020,Borgnia2020,ZSYang2020,Yokomizo2020},
and the exceptional points (EPs) \cite{Heiss2012,Miri2019,BergholtzKunst2021},
the non-Hermitian physics has been attracting increasing attention.
Many non-Hermitian phenomena have been explored in various quantum platforms,
including quantum optics \cite{LXiao2017,Weidemann2020,KWang2021},
quantum spin systems \cite{YWu2019,WQLiu2020,WGZhang2020},
ultracold atoms \cite{DWZhang2018,JMLi2019,WGou2020,LZTang2020,DWZhang2020a,LZTang2021,LFZhang2021,LZTang2022,ZJRen2022}, and so on.
However, none of these studies has discussed the direct measurement of the non-Hermitian generalization of the expectation value, $\bra{\psi^L} A \ket{\psi^R}$, with $A$ being an arbitrary operator and $\ket{\psi^{L,R}}$ being a pair of left and right eigenstates, dubbed {\it dual eigenstates}, of a given non-Hermitian Hamiltonian, which is deeply involved in many definitions of non-Hermitian quantities as the straightforward generalization of the Hermitian counterparts \cite{Moiseyev2011}.
Developing a method of measuring quantities of this form is urgent and may be a prerequisite for studying in a universal manner the exotic non-Hermitian phenomena in experiments.

One of the most interesting quantities involving $\bra{\psi^L} A \ket{\psi^R}$ is the non-Hermitian topological invariant \cite{Ashida2020}, such as the non-Hermitian winding number \cite{CHYin2018}.
For example, the Su-Schrieffer-Heeger (SSH) model \cite{SuHeeger1979} with nonreciprocal hopping is a prototypical non-Hermitian topological model; the difference between it and its Hermitian counterpart is reflected by the winding number defined with the dual eigenstates \cite{Moiseyev2011}.
Existing works try to establish relations between this non-Hermitian winding number and the experimentally measurable expectation values to figure it out indirectly.
It was shown that the winding number can be calculated by the dynamic winding numbers, defined by the integral of the long-time average of measurable expectation values \cite{BZhu2020}.
On the other hand, the authors of Ref. \cite{WGZhang2020} reparametrize the non-Hermitian Hamiltonian and use the measurable expectation values to fit the parameters; the winding number is reconstructed by the parameters.
The limitation of these works is the lack of generality for measuring the quantity $\bra{\psi^L} A \ket{\psi^R}$, and the relations they found may just be available in special cases.

Furthermore, a quantity such as $\bra{\psi^L} A \ket{\psi^R}$ is not only restricted within the non-Hermitian physics if the dual eigenstates are relaxed to two arbitrary quantum states $| \psi_{1,2} \rangle$, i.e., $\langle \psi_1 | A | \psi_2 \rangle$, which we dub the {\it generalized expectation} of $A$ in the following as a two-state generalization of the conventional expectation.
Quantities in the form of a generalized expectation are ubiquitous in quantum physics, endowed with distinct meanings in different fields of physics \cite{Sakurai2017}, such as
the overlap integral of two states,
matrix elements of an operator,
scattering amplitudes, and the Green's function or Feynman propagator in quantum field theory.
Since direct measurement of them is outside of the conventional formalism, e.g., the projective (von Neumann) measurement \cite{Nielsen2010}, some indirect methods have been proposed, such as the {\textsc{swap}} test \cite{Buhrman2001}, weak measurement \cite{Dressel2014}, and quantum circuits for simulating the correlation function \cite{Ortiz2001,Pedernales2014,Francis2020}.
However, these proposals are still only valid for specific situations; that is, $|\psi_{1,2}\rangle$ cannot be arbitrary.
A general scheme for measuring generalized expectations remains to be settled even in broader fields of quantum physics.

To deal with the measurement of $\bra{\psi_1} A \ket{\psi_2}$, we propose a general circuit based on the {\textsc{swap}} test \cite{Buhrman2001} in Fig. \ref{fig: measure circuit} to directly capture the real and imaginary parts of the generalized expectation.
Meanwhile, to apply the general circuit to $\bra{\psi^L} A \ket{\psi^R}$ in non-Hermitian systems, a quantum circuit (Fig. \ref{fig: prepare circuit with dilation method}) for efficiently preparing the dual eigenstates of a given non-Hermitian Hamiltonian as the input of the general circuit is also developed with the aid of the dilation method \cite{YWu2019}.
By numerically simulating these circuits in the nonreciprocal SSH model, we successfully obtain the Bloch and non-Bloch spin textures and the corresponding winding numbers under periodic and open boundary conditions (PBCs and OBCs), respectively, which demonstrates the validity of our circuits.

The paper is organized as follows.
The general quantum circuit for measuring generalized expectations is proposed in Sec. \ref{sec: quantum circuit}.
Specially for non-Hermitian systems, the quantum circuit for preparing the dual eigenstates of a non-Hermitian Hamiltonian is developed in Sec. \ref{sec: NH bi prep}.
In Sec. \ref{sec: applications}, we apply these circuits to the non-reciprocal SSH model and numerically simulate the measurement of Bloch and non-Bloch spin textures and the winding numbers.
Section \ref{sec: conclusion} provides a conclusion.

\section{A general measurment circuit for generalized expectations} \label{sec: quantum circuit}

Our aim is to measure the quantity, $\langle \psi_1|A|\psi_2\rangle$, of an arbitrary operator $A$ with respect to two quantum states $\ket{\psi_1}$ and $\ket{\psi_2}$, dubbed a {\it generalized expectation} of $A$, which  reduces to the conventional expectation when the two states are identical, i.e., $\ket{\psi_1}=\ket{\psi_2}$.
Because any operator can be decomposed into Hermitian operators, $A=\left(\frac{A+A^\dag}{2}\right)+i\left(\frac{A-A^\dag}{2i}\right)$, we just need to propose a quantum circuit to measure the generalized expectation of a {\it Hermitian} operator, i.e., $\langle \psi_1|O|\psi_2\rangle$, where $O$ represents an experimentally accessible, Hermitian operator.

\begin{figure}[tb]
    \includegraphics[]{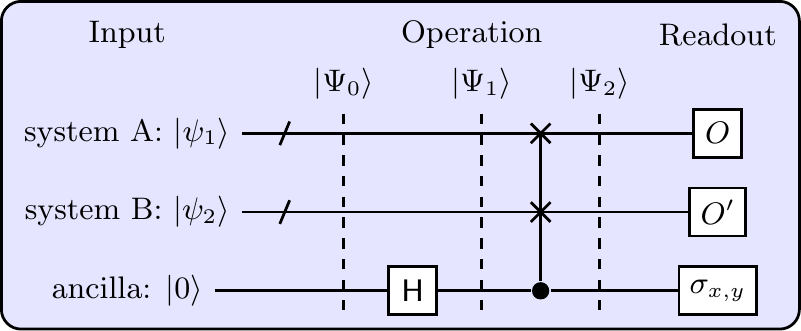}
	\caption{The general quantum circuit for measuring the generalized expectation $\braket{\psi_1|O|\psi_2}$ scaled by $\braket{\psi_1|O'|\psi_2}$.
		Two quantum states $|\psi_{1,2}\rangle$ are assumed to be prepared by other methods as input in systems A and B, respectively, each of which consists of $n$ qubits. The ancilla qubit is initialized to $|0\rangle$.
			The successive operations include a Hadamard gate (denoted by $\mathsf{H}$) on the ancilla and a controlled-{\textsc{swap}} {(Fredkin)} gate with the ancilla as the control qubit.
			$O$ and $O'$ denote the experimentally accessible Hermitian operators. $\sigma_{x,y}$ are two operators of Pauli matrices. The details of the readout process are given in Appendix \ref{sec: measurement details}.}
    \label{fig: measure circuit}
\end{figure}

Figure \ref{fig: measure circuit} shows the quantum circuit for measuring $\braket{\psi_1|O|\psi_2}$, which is the main result of this paper. Supposing that $| \psi_{1,2} \rangle$ are obtained,  this circuit consists of systems A and B, each of which is represented by $n$ qubits, and an ancilla qubit.
Firstly, the two states $|\psi_{1,2}\rangle$ are put into systems A and B, respectively, and the ancilla qubit is initialized to $|0\rangle$, yielding a product state
\begin{equation}
    \ket{\Psi_0}=\ket{\psi_1} \otimes \ket{\psi_2} \otimes \ket{0}
\end{equation}
as an initial state.

Secondly, by applying a Hadamard gate to the ancilla, and then a controlled-{\textsc{swap}} (Fredkin) gate with the ancilla being the control qubit, we obtain successively
\begin{align}
    \ket{\Psi_1}=&\ket{\psi_1} \otimes \ket{\psi_2} \otimes \frac{1}{\sqrt{2}}(\ket{0}+\ket{1}), \label{eq: Psi 1}\\
    \ket{\Psi_2}=&\frac{1}{\sqrt{2}} \left( \ket{\psi_1} \otimes \ket{\psi_2} \otimes \ket{0} +\ket{\psi_2} \otimes \ket{\psi_1} \otimes \ket{1} \right) . \label{eq: Psi 2}
\end{align}

Finally, the operator $O$ and an ancillary Hermitian operator $O'$ are introduced to systems A and B, respectively.
Because of the Hermiticity of $O$ and $O'$, we obtain the following relations (see Appendix \ref{sec: details of eqs: expectation} for the detailed derivation):
\begin{equation} \label{eqs: expectation}
    \begin{split}
        \frac{\braket{\Psi_2|O \otimes O' \otimes \sigma_x|\Psi_2}}{\braket{\Psi_2|O' \otimes O' \otimes \sigma_x|\Psi_2}} =& \,\Re \left( \frac{\braket{\psi_1|O|\psi_2}}{\braket{\psi_1|O'|\psi_2}} \right), \\
        \frac{\braket{\Psi_2|O \otimes O' \otimes \sigma_y|\Psi_2}}{\braket{\Psi_2|O' \otimes O' \otimes \sigma_x|\Psi_2}} =& \,\Im \left( \frac{\braket{\psi_1|O|\psi_2}}{\braket{\psi_1|O'|\psi_2}} \right),
    \end{split}
\end{equation}
where $\sigma_{x,y}$ are two operators of Pauli matrices.
Assuming that $O$ and $O'$ are experimentally accessible, from Eq. \eqref{eqs: expectation} the generalized expectation $\braket{\psi_1|O|\psi_2}$ scaled by $\braket{\psi_1|O'|\psi_2}\ne 0$ can be figured out by measuring the traditional expectations [left-hand side of Eq. \eqref{eqs: expectation}] in experiment.
Based on current experiment scales \cite{Google2023}, this general measurement quantum circuit in Fig. \ref{fig: measure circuit} can be applied to the qubit systems with states and operators being composed of up to $\sim\! 50$ qubits of each.
See Appendix \ref{sec: measurement details} for details of the measurement.

For convenience, $O'$ can be set as an identity operator, and the scaling factor $\braket{\psi_1|O'|\psi_2}$ reduces to $\braket{\psi_1|\psi_2}$, which is usually not equal to $0$. Sometimes, if the two states are orthogonal, i.e., $\braket{\psi_1|\psi_2}=0$, $O'$ should be properly selected such that $\braket{\psi_1|O'|\psi_2} \neq 0$, and the generalized expectation with respect to these two orthogonal states can also be measured only up to a constant.

\section{Preparation for dual eigenstates of a non-Hermitian Hamiltonian} \label{sec: NH bi prep}

For different purposes, we are interested in various quantities with the form of a generalized expectation.
In the context of non-Hermitian physics, the generalized expectation of a Hermitian operator $O$ with respect to a pair of dual eigenstates  $|\psi^{R,L}\rangle$ is defined as
\begin{equation} \label{eq: NH expectation}
    \braket{O}_{\mathrm{NH}} \equiv \frac{\braket{\psi^L|O|\psi^R}}{\braket{\psi^L|\psi^R}},
\end{equation}
which usually emerges to characterize important non-Hermitian quantities. Here, $|\psi^{R}\rangle$ is one right eigenstate of a given non-Hermitian Hamiltonian $H$ with the eigenenergy $E$,  and $\langle\psi^{L}|$ is the corresponding left eigenvector of $H$, i.e., $H|\psi^R\rangle = E |\psi^R\rangle$ and $H^\dagger|\psi^L\rangle = E^* |\psi^L\rangle$; $|\psi^{R,L}\rangle$ form a pair of dual eigenstates of $H$.
To measure $\braket{O}_{\mathrm{NH}}$, according to  Eq. (\ref{eqs: expectation}), we just need to prepare $|\psi_{1,2}\rangle = |\psi^{L,R}\rangle$ as the input states of the general circuit in Fig. \ref{fig: measure circuit} and $O'$ is set as the identity operator.

Contrary to the adiabatic-evolution method of generating eigenstates of a given Hermitian Hamiltonian, the amplifying and decaying feature, imprinted in complex eigenenergies, of non-Hermitian Hamiltonians offers a more convenient principle to prepare eigenstates \cite{HLWang2018,WGZhang2020}:
If some eigenenergies of a non-Hermitian Hamiltonian $H$ have nonvanishing imaginary parts, the long-time nonunitary evolution will lead an arbitrary initial state that is a superposition of these eigenstates to the state composed of the eigenstates with the largest imaginary part of the eigenenergies.
Therefore, with this principle, the dual eigenstates $\ket{\psi^R}$ and $\ket{\psi^L}$ should be obtained by respective evolutions under $H$ and $-H^\dagger$, of which the eigenvalues, $E$ and $-E^*$, can have the largest imaginary parts simultaneously.
In the same spirit, when the eigenenergies of the target dual eigenstates are purely real, we are able to use the Hamiltonian pair $(\alpha H,-\alpha^*H^\dag)$ with a complex multiplier $\alpha$ effectively rotating the original Hamiltonians in the complex plane to have good control over the evolution.

\begin{figure}[tb]
    \includegraphics[width=0.98\columnwidth]{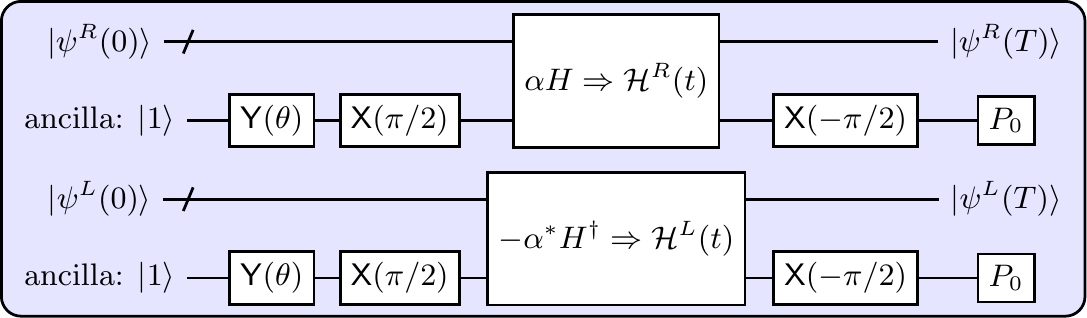}
    \caption{The quantum circuit of preparing a pair of dual eigenstates of a given non-Hermitian Hamiltonian $H$, as the input of the general circuit in Fig. \ref{fig: measure circuit}. The target right and left eigenstates of $H$ can be reached by long-time evolutions under $\alpha H$ and $-\alpha^*H^\dag$, respectively. The selection of the complex multiplier $\alpha$ depends on the specific case.
        $\mathcal{H}^{R,L}(t)$ are dilated Hermitian Hamiltonians in the dilation method, determining $\alpha H$ and $-\alpha^* H^\dagger$, respectively. $P_0 = | 0 \rangle \langle 0 |$ is a projection operator on state $|0\rangle$ in the ancillas for postselection. Other notations are explained in the main text.}
    \label{fig: prepare circuit with dilation method}
\end{figure}

To experimentally implement a quantum evolution of a non-Hermitian system, several methods are proposed \cite{YWu2019,WQLiu2020,DJZhang2019b,JWWen2019,GLZhang2020}. Here, we use the dilation method developed by Wu and co-workers \cite{YWu2019,WQLiu2020} with the demanded $(\alpha H,-\alpha^*H^\dag)$ being the target effective non-Hermitian Hamiltonian pair (see Appendix \ref{sec: dilation method} for details of the dilation method).
Figure \ref{fig: prepare circuit with dilation method} sketches the quantum circuit for preparing a pair of dual eigenstates of a given non-Hermitian Hamiltonian $H$.
Two initial states $\ket{\psi^{R,L}(t=0)}$ are input to the circuits with both ancillas being set to $\ket{1}$.
After successively applying $\mathsf{Y}(\theta)$ and $\mathsf{X}(\pi/2)$, which are qubit rotations about the $y$ and $x$ axes by angles $\theta=2\arctan\eta_0$ (where $\eta_0$ is a parameter of the dilation method, depending on specific Hamiltonians; see Appendix \ref{sec: dilation method} for details) and $\pi/2$, to the ancillas, the qubits evolve under the dilated Hermitian Hamiltonian $\mathcal{H}^{R,L}(t)$, derived by $\alpha H$ and $-\alpha^* H^\dagger$, respectively. With $\mathsf{X}(-\pi/2)$ being applied to each ancilla at the end of the evolution, the final states $\ket{\psi^{R,L}(t=T\rightarrow\infty)}\rightarrow|\psi^{R,L} \rangle$, where $T$ represents the evolution time, are just the demanded pair of dual eigenstates when the ancillas are measured to be $\ket{0}$.
In principle, any nondegenerate eigenstates can be prepared by this method.

\section{Applications to non-Hermitian winding numbers} \label{sec: applications}

In this section, we take the pedagogical SSH model with nonreciprocal hopping \cite{Lee2016,CHYin2018,SYYao2018a} as an example, and apply our proposed circuits of Figs. \ref{fig: measure circuit} and \ref{fig: prepare circuit with dilation method} to the measurement of non-Hermitian spin textures and winding numbers, including Bloch spin textures and winding numbers under PBCs and non-Bloch spin textures and winding numbers under OBCs, which are much different from their Hermitian counterparts.

\subsection{Bloch winding numbers} \label{sec: bloch winding}

The Hamiltonian of the nonreciprocal SSH model under PBCs in $k$ space reads
\begin{equation}\label{eq: Hamiltonian}
    H(k)=\begin{bmatrix}
        0 & t_1- \delta+t_2 e^{-ik}\\
        t_1+ \delta+t_2 e^{ik}& 0
    \end{bmatrix}
\end{equation}
where $t_1\pm\delta$ are the nonreciprocal intra-cell hopping amplitudes and $t_2$ is the reciprocal inter-cell hopping amplitude; all the parameters are real.
In terms of Pauli matrices, $H(k)= \mathbf{d}(k)\cdot \boldsymbol{\sigma}$, where $\boldsymbol{\sigma}=(\sigma_x,\sigma_y,\sigma_z)$ is a vector of Pauli matrices and $\mathbf{d}(k)=[d_x(k),d_y(k),d_z(k)]=(t_1+t_2 \cos k,t_2 \sin k -i \delta,0)$ can be regarded as an effective complex magnetic field.

The two eigenenergies of $H(k)$ read $E_\pm(k) = \pm d(k)=\pm\sqrt{d^2_x(k)+d^2_y(k)+d^2_z(k)}$ and the corresponding right and left eigenstates are $\ket{\psi^{R,L}_\pm(k)}$.
With the pair of dual eigenstates labeled by $+$, the Bloch spin texture in this non-Hermitian model can be defined as
\begin{equation} \label{eq: spin texture}
    \mathbf{n}(k) \equiv \braket{\boldsymbol{\sigma}(k)}_{\mathrm{NH}} = \frac{\braket{\psi^L_+(k)|\boldsymbol{\sigma}|\psi^R_+(k)}}{\braket{\psi^L_+(k)|\psi^R_+(k)}},
\end{equation}
% which is also complex in general.
which can also be expressed in terms of another pair of dual eigenstates $|\psi_-^{R,L}(k)\rangle$.
This formula follows the structure of the generalized expectation, so we can measure it using our proposed measurement circuit of Fig. \ref{fig: measure circuit} with the replacement of $O$ by $\boldsymbol{\sigma}$ and $O'$ by the identity matrix in Eq. \eqref{eqs: expectation}, as long as the dual eigenstates $|\psi_+^{R,L}(k)\rangle$ as the input are well prepared, and $\braket{\psi^L_+(k)|\psi^R_+(k)}\ne 0$, which excludes the EPs; the cases for EPs and points nearby are discussed at the end of this section.

For the non-Hermitian model \eqref{eq: Hamiltonian}, the pair of dual eigenstates $\ket{\psi^{R,L}_+(k)}$ at each $k$ can be obtained following the recipe in Sec. \ref{sec: NH bi prep} by evolving an initial state $\ket{\psi^{R,L}(t=0)}=\ket{0}$ under $H(k)$ and $-H^\dag(k)$ ($\alpha=1$), respectively, if $\Im\{E_+(k)\}>\Im\{E_-(k)\}$, and otherwise, under $-H(k)$ and $H^\dag(k)$ ($\alpha=-1$), respective, if $\Im\{E_+(k)\}<\Im\{E_-(k)\}$;
when $\Im\{E_+(k)\}$ is close to $0$, the measurement accuracy of the spin texture $\mathbf{n}(k)$ can be improved by adjusting $\alpha$ to ensure $\Im\{\alpha E_+(k)\}>\Im\{\alpha E_-(k)\}$.
With the prepared dual eigenstates, the general circuit of Fig. \ref{fig: measure circuit} is applied;  the details of the readout process are given in Appendix \ref{sec: measurement details}.
Thus, according to Eq. \eqref{eqs: expectation}, the non-Hermitian spin textures can be obtained.

With the measured spin texture $\mathbf{n}(k)$, the non-Hermitian Bloch winding number, defined as \cite{CHYin2018}
\begin{equation} \label{eq: winding number}
    w_\mathrm{B} = \frac{1}{2\pi} \int_{-\pi}^\pi \partial_k \phi(k) ~ dk,
\end{equation}
can be recast by the complex angle $\phi(k)\equiv\tan^{-1}[d_y(k)/d_x(k)]=\tan^{-1}[n_y(k)/n_x(k)]$.
For the Hermitian case ($\delta=0$), this winding number classifies a topological phase with $w_\mathrm{B}=1$ and a topologically trivial phase with $w_\mathrm{B}=0$ \cite{SQShen2012}.
However, the non-Hermitian case ($\delta \neq 0$) has three topologically distinct phases \cite{CHYin2018}:
(1) $w_\mathrm{B}=1$, for $|t_1|+|\delta|<|t_2|$;
(2) $w_\mathrm{B}=1/2$, for $\left||t_1|-|\delta|\right|<|t_2|<|t_1|+|\delta|$, which has no counterpart in Hermitian systems;
and (3) $w_\mathrm{B}=0$, for $\left||t_1|-|\delta|\right|>|t_2|$.

\begin{figure}[tb]
    \centering
    \includegraphics[width=\columnwidth]{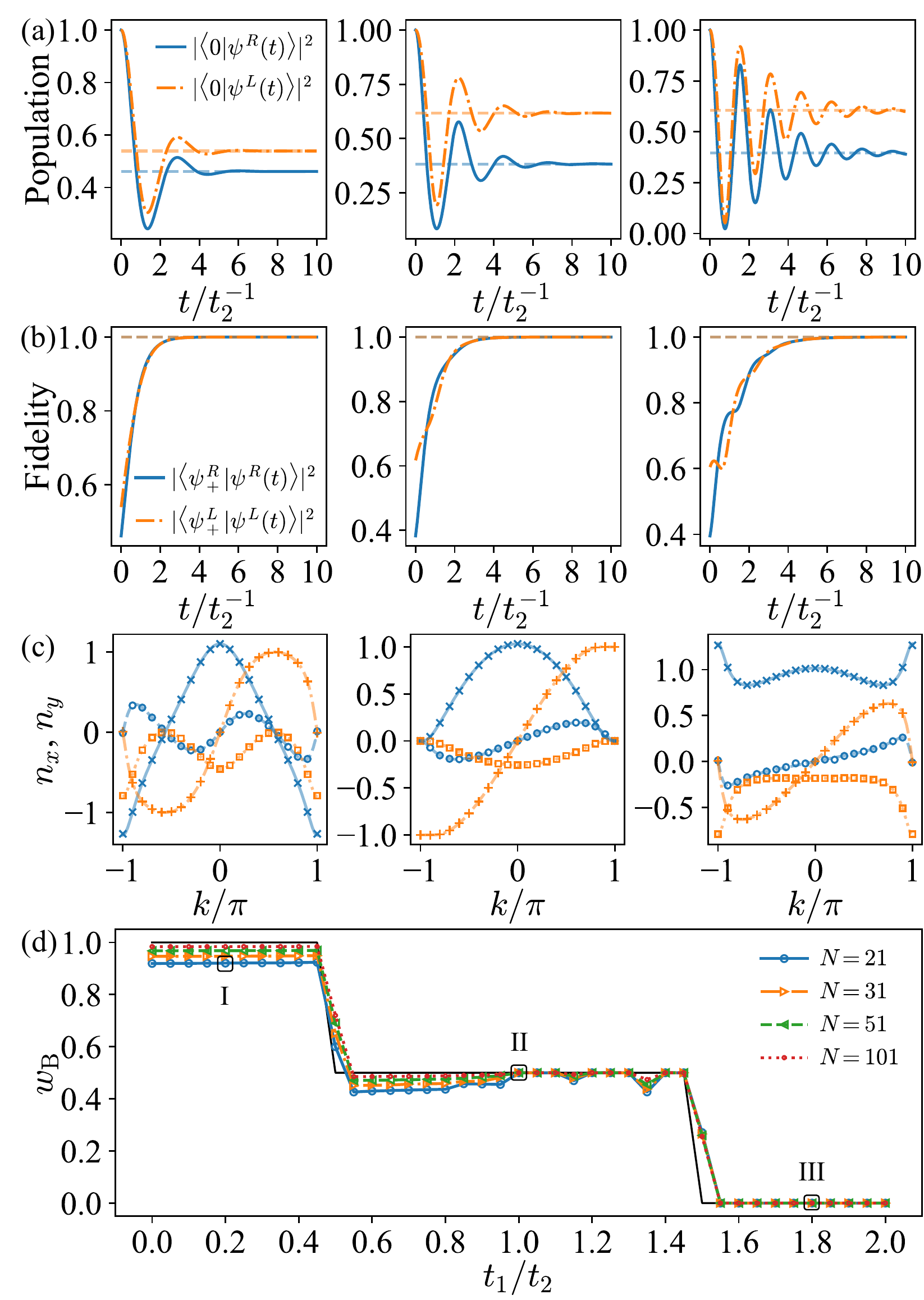}
    \caption{(a) and (b) Evolution of the population and the fidelity at $k=\pi/2$, both of which converge to the demanded pair of dual eigenstates (dashed lines).
        The left ($t_1/t_2 = 0.2$), middle ($t_1/t_2 = 1$), and right ($t_1/t_2 = 1.8$) panels correspond to the circled points I, II, and III in (d), respectively.
        (c) The spin textures generated by the analytical calculation (lines) and the numerical circuit simulation (symbols).
        The blue and orange colors denote $n_{x}(k)$ and $n_{y}(k)$, respectively; 
        the solid (dashed) lines and the crosses (circles) denote the real (imaginary) parts of $n_{x}(k)$, and the dash-dotted (dotted) lines and the pluses (squares) denote the real (imaginary) parts of $n_{y}(k)$.
        (d) The Bloch winding numbers with different sampling numbers $N$.
        The data in (c) and (d) are taken at the evolution time  $T=10/t_2$. The parameter $\delta/t_2 = 0.5$ for all figures.
    }
    \label{fig: winding number}
\end{figure}

To demonstrate the validity of our scheme and the accuracy of the preparation for dual eigenstates, we numerically simulate our quantum circuits and calculate the Bloch winding numbers. The results are shown in Fig. \ref{fig: winding number}.
When the evolution time takes the order of $T=10/t_2$, the dual eigenstates $\ket{\psi^{R,L}_+}$ are well captured by the final states $\ket{\psi^{R,L}(t=T)}$ with high accuracies in terms of population [Fig. \ref{fig: winding number}(a)] and with high fidelities [Fig. \ref{fig: winding number}(b)].
The spin textures $n_{x,y}(k)$ calculated using the final states coincide very well with the analytical results [Fig. \ref{fig: winding number}(c)], and thus the Bloch winding numbers $w_\mathrm{B}$ approach the expected values when the sampling number $N$ becomes larger and larger [Fig. \ref{fig: winding number}(d)].
The topological phase transitions are also reflected near the points $t_1/t_2=0.5$ and $1.5$.

Here, we directly perform the nonunitary evolutions in our numerical calculations without using the dilation method, because the dilated Hamiltonians $\mathcal{H}^{R,L}[k(t)]$ need be fine-tuned case by case for thousands of sampling points, but the validity of the dilation method is demonstrated for some instances in Appendix \ref{sec: dilation method}.
Because the measurement up to now does not involve EPs except the transition points, we can safely remove $O'$ in Eq. \eqref{eqs: expectation} due to $\langle\psi^L_n|\psi^R_n\rangle\ne0$.
At or near EPs, the orthogonality of dual eigenstates may fail the measurement due to the vanishing of the denominator in Eq. \eqref{eq: NH expectation}.
For these scenarios, one way is to find a simple operator $O'$ such that $\langle\psi^L_n|O'|\psi^R_n\rangle\ne0$ and the measured spin texture is the same up to a scaling factor; the other way is to set $O$ and $O'$ as $\sigma_y$ and $\sigma_x$, respectively, to directly measure the angle $\phi(k)$ determined by the ratio $n_y(k)/n_x(k)$ that bypasses the orthogonal condition.

\subsection{Non-Bloch winding numbers} \label{sec: non-bloch winding}

The existence of topological edge states localized at the ends of the non-Hermitian SSH chain under OBCs cannot be correctly predicted by the Bloch winding number calculated under PBCs, which is the phenomenon of breakdown of the bulk-boundary correspondence in non-Hermitian systems \cite{Lee2016,SYYao2018a,MartinezAlvarez2018a,YXiong2018}. This is due to the non-Hermitian skin effect \cite{SYYao2018b} that has no Hermitian counterpart.
Yao and Wang \cite{SYYao2018a} proposed a non-Bloch winding number defined in the generalized Brillouin zone (GBZ) under OBCs, successfully restoring the correspondence.
Using our proposed circuit, we can also measure this important non-Hermitian quantity.

The generalized Brillouin zone is defined by a complex variable $\beta$ with a modulus $|\beta| \equiv r \ne 1$ in general.
After replacing $e^{ik}$ in Eq. \eqref{eq: Hamiltonian} by $\beta$, the non-Bloch version of the Hamiltonian reads
\begin{equation}\label{eq: non-Bloch Hamiltonian}
    H(\beta)=\begin{bmatrix}
        0 & t_1-\delta+\beta^{-1} t_2 \\
        t_1+\delta+\beta t_2 & 0
    \end{bmatrix}.
\end{equation}
For this nonreciprocal SSH model, we can make the parametrization $\beta\equiv re^{ik}$ with $r = \sqrt{\left|\frac{t_1+\delta}{t_1-\delta}\right|}$ and $k$ being a real parameter that is similar to the Bloch wave number.  Thus $\beta$ can take values in a nonunit circle in the complex plane.
Like the Bloch Hamiltonian, the non-Bloch Hamiltonian \eqref{eq: non-Bloch Hamiltonian} can also be written as $H(\beta) = \mathbf{d}(\beta) \cdot \boldsymbol{\sigma}$, where $\mathbf{d}(\beta) = [d_x(\beta), d_y(\beta), d_z(\beta)] = (t_1+\frac{\beta+\beta^{-1}}{2} t_2, \frac{\beta-\beta^{-1}}{2i} t_2 -i \delta, 0)$.
The non-Bloch winding number has the same form as the Bloch one, yielding
\begin{equation} \label{eq: non-Bloch winding number}
    w_\mathrm{N} = \frac{1}{2\pi} \int_{C_\beta} \partial_\beta \phi(\beta) ~ d\beta = \frac{1}{2\pi} \int_{-\pi}^\pi \partial_k \phi[\beta(k)] ~ dk ,
\end{equation}
where $\phi(\beta)\equiv\tan^{-1}[d_y(\beta)/d_x(\beta)]=\tan^{-1}[n_y(\beta)/n_x(\beta)]$ is determined by the non-Bloch spin texture $\mathbf{n}(\beta)$ that is of the same form as Eq. \eqref{eq: spin texture} with only the replacement of $k$ by $\beta$.
In view of this non-Bloch winding number, this nonreciprocal SSH model has two topologically distinct phases: $w_\mathrm{N} = 1$ for $|t_1^2-\delta^2|<t_2^2$, which supports topological edge states, and otherwise, $w_\mathrm{N}=0$, where there is no edge state.

\begin{figure}[tb]
    \includegraphics[width=\columnwidth]{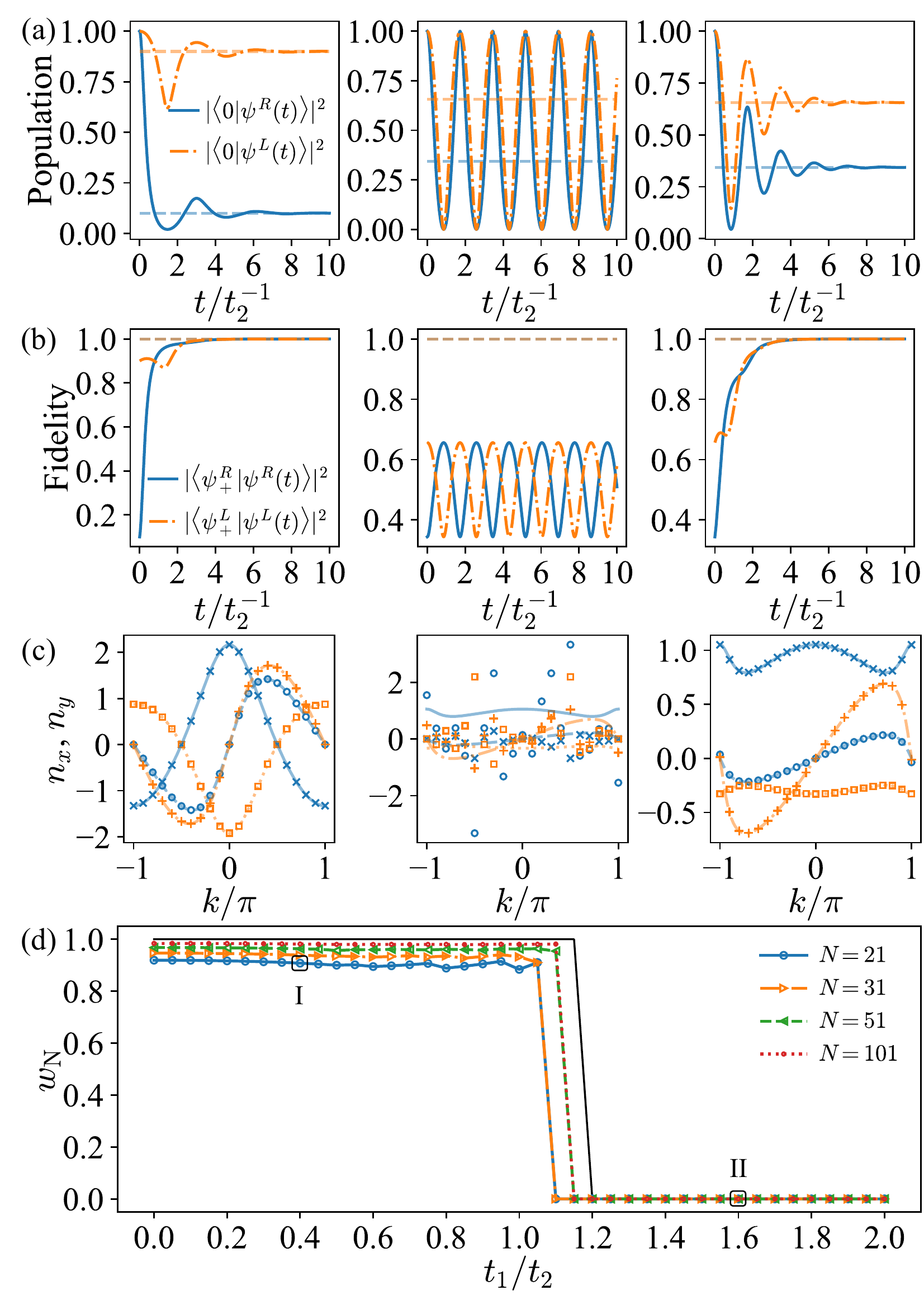}
    \caption{(a) and (b) Evolution of the population and the fidelity at $k=\pi/2$. The left panel ($t_1/t_2 = 0.4$) corresponds to the circled point I in (d), and the middle and right panels ($t_1/t_2 = 1.6$) both correspond to the circled point II in (d).
    Since the model has a purely real spectrum when $t_1/t_2=1.6$, we demonstrate the validity of preparation method in Sec. \ref{sec: NH bi prep} with $\alpha = 1$ and $e^{i\pi/16}$ in the middle and right panels, respectively.
    (c) The spin textures and (d) the non-Bloch winding numbers; the lines and symbols in (c) have the same meaning as in Fig. \ref{fig: winding number}(c).
    Other parameters are also the same as those in Fig. \ref{fig: winding number}.}
    \label{fig: nb winding numbers}
\end{figure}

Similar circuits are used as in the previous Bloch part for preparing dual eigenstates $|\psi_+^{R,L}(\beta)\rangle$ and measuring the spin textures $ \mathbf{n}(\beta)$, and then, non-Bloch winding numbers can be recast likewise.
The simulation results with high accuracy compared with the analytical ones are shown in Fig. \ref{fig: nb winding numbers} using the same system parameters as those in Fig. \ref{fig: winding number}.
The main difference comes from the reality of the spectrum at certain parameters (e.g., $t_1/t_2=1.6$), and a proper multiplier $\alpha$ should be introduced to make the preparation of dual eigenstates more efficient, which can be seen by comparing the middle and right panels of Figs. \ref{fig: nb winding numbers}(a) and \ref{fig: nb winding numbers}(b).
In Fig. \ref{fig: nb winding numbers}(d), one can see the topological phase transition near $t_1/t_2=1.2$, different from the characterization of Bloch winding numbers.

\section{Conclusion} \label{sec: conclusion}

We propose a quantum circuit (Fig. \ref{fig: measure circuit}) for measuring an expectationlike quantity, $\langle \psi_1|O|\psi_2\rangle$, of a Hermitian operator $O$ with respect to two given quantum states $\ket{\psi_1}$ and $\ket{\psi_2}$, and thus the general quantity, $\langle \psi_1|A|\psi_2\rangle$, dubbed the generalized expectation, of an arbitrary operator $A$.
With the aid of the operators $\sigma_x$ and $\sigma_y$ acting on the ancilla of the circuit, both real and imaginary parts of the generalized expectation can be obtained via the experimentally accessible conventional expectations, i.e., the left-hand side of Eq. (\ref{eqs: expectation}), which are the main results of this paper.

Then, we apply the general circuit to the measurement of generalized expectations in non-Hermitian systems, $\braket{O}_{\mathrm{NH}}$, where dual eigenstates of non-Hermitian Hamiltonians are usually used; therefore we also propose an efficient circuit (Fig. \ref{fig: prepare circuit with dilation method}), in light of the dilation method \cite{YWu2019,WQLiu2020}, to prepare the dual eigenstates by effectively rotating the Hamiltonian in the complex plane.
As an application, we take the nonreciprocal SSH model as an example and numerically simulate the measurement using the proposed circuits, obtaining the Bloch and non-Bloch spin textures and the corresponding winding numbers under PBCs and OBCs, respectively. The results are as good with high fidelity as expected by the theories (Figs. \ref{fig: winding number} and \ref{fig: nb winding numbers}).

In principle, other non-Hermitian topological invariants composed of non-Hermitian spin textures, e.g., non-Hermitian Chern numbers \cite{Fukui2005,DLDeng2014} and Wilson loops \cite{Tracy2016}, can also be measured following our schemes.
In the broader settings of physics, the specific meaning of the generalized expectation, $\langle \psi_1|A|\psi_2\rangle$, such as the overlap integral, correlation function, scattering amplitude, etc., endows our circuits with more potential applications.

\appendix

\section{Derivation of Eq. \eqref{eqs: expectation}} \label{sec: details of eqs: expectation}
In this appendix, we give the details of derivation of Eq. \eqref{eqs: expectation} in the main text.

Substituting $\ket{\Psi_2}$ of Eq. \eqref{eq: Psi 2} in the main text into the following quantities,
we have
\begin{eqnarray}
	&&\braket{\Psi_2|O \otimes O' \otimes \sigma_x|\Psi_2} \notag \\
	&=& \frac{1}{2}\left(\braket{\psi_1|O|\psi_2}\braket{\psi_2|O'|\psi_1} + \braket{\psi_2|O|\psi_1}\braket{\psi_1|O'|\psi_2} \right) \notag \\
	&=& \Re (\braket{\psi_1|O|\psi_2}\braket{\psi_2|O'|\psi_1}), \\
	&&\braket{\Psi_2|O \otimes O' \otimes \sigma_y|\Psi_2} \notag \\
	&=& \frac{1}{2i}\left(\braket{\psi_1|O|\psi_2}\braket{\psi_2|O'|\psi_1} - \braket{\psi_2|O|\psi_1}\braket{\psi_1|O'|\psi_2} \right) \notag \\
	&=& \Im (\braket{\psi_1|O|\psi_2}\braket{\psi_2|O'|\psi_1}).
\end{eqnarray}
When $O=O'$, it is found that
\begin{eqnarray}
	\braket{\Psi_2|O' \otimes O' \otimes \sigma_x|\Psi_2} &=& |\langle\psi_1|O'|\psi_2\rangle|^2 \ge 0,\\
	\braket{\Psi_2|O' \otimes O' \otimes \sigma_y|\Psi_2} &=& 0.
\end{eqnarray}
We can choose $O'$ such that $|\langle\psi_1|O'|\psi_2\rangle|^2 \ne 0$, and then Eq. \eqref{eqs: expectation} is obtained as follows:
\begin{widetext}
	\begin{equation}
    \begin{split}
        \frac{\braket{\Psi_2|O \otimes O' \otimes \sigma_x|\Psi_2}}{\braket{\Psi_2|O' \otimes O' \otimes \sigma_x|\Psi_2}} &= \frac{\Re[\langle \psi_1 | O | \psi_2 \rangle \langle \psi_2 | O' | \psi_1 \rangle]}{|\langle \psi_1 | O' | \psi_2 \rangle|^2} = \Re \left( \frac{\braket{\psi_1|O|\psi_2}}{\braket{\psi_1|O'|\psi_2}} \right), \\
        \frac{\braket{\Psi_2|O \otimes O' \otimes \sigma_y|\Psi_2}}{\braket{\Psi_2|O' \otimes O' \otimes \sigma_x|\Psi_2}} &= \frac{\Im[\langle \psi_1 | O | \psi_2 \rangle \langle \psi_2 | O' | \psi_1 \rangle]}{|\langle \psi_1 | O' | \psi_2 \rangle|^2} = \Im \left( \frac{\braket{\psi_1|O|\psi_2}}{\braket{\psi_1|O'|\psi_2}} \right).
    \end{split}
\end{equation}
\end{widetext}

\section{Details of measurement} \label{sec: measurement details}
In this appendix, we demonstrate how to measure $\braket{\Psi_2|O \otimes O' \otimes \sigma_{x,y}|\Psi_2}$, which is the target observable of the general circuit of Fig. \ref{fig: measure circuit} in the main text.

Define the projection operators $P_{0} = \left|0\right\rangle\! \left\langle 0 \right|$ and $P_{1} = \left|1\right\rangle \!\left\langle 1\right|$, where $|0\rangle$ and $|1\rangle$ are the two eigenstates of Pauli operator $\sigma_z$.
The expectation of $\sigma_z$ can be written as
\begin{eqnarray}
\langle\sigma_z\rangle&\equiv&\left\langle \psi \left| \sigma_z \right| \psi \right\rangle = \langle \psi | P_{0} - P_{1} | \psi \rangle\notag\\
&=&|\braket{0|\psi}|^2-|\braket{1|\psi}|^2 \approx  (N_{0} - N_{1})/N,
\end{eqnarray}
where we use $N_{s}~(s=0,1)$ to represent the number of times the eigenstate $|s\rangle$ is detected when measuring the state $|\psi\rangle$, and $N = N_0+N_1$ is the total number of measurement times.
This equation means that the expectation of $\sigma_z$ can be measured by counting $N_0$ and $N_1$.

An arbitrary Hermitian operator acting on a single qubit can be written as $\mathbf{d} \cdot \boldsymbol{\sigma}+d_0\sigma_0$, where $\mathbf{d}=(d_x,d_y,d_z)$ is a real-valued vector with the norm defined by $d$, $\boldsymbol{\sigma}=(\sigma_x,\sigma_y,\sigma_z)$ is the vector of Pauli operators, and $\sigma_0$ represents the identity operator.
With the help of a counterclockwise rotation $\mathsf{U}(\theta)=e^{-i \frac{\theta}{2} \hat{\mathbf{u}} \cdot \boldsymbol{\sigma}}$ of a qubit by an angle $\theta = \arccos(\hat{\mathbf{d}} \cdot \hat{\mathbf{z}})\in[0,\pi]$ about the axis $\hat{\mathbf{u}} \equiv \hat{\mathbf{d}} \times \hat{\mathbf{z}}/\sin\theta$, the Hermitian operator can be rewritten as $\mathbf{d} \cdot \boldsymbol{\sigma}+d_0\sigma_0= \mathsf{U}^{\dagger}(d\sigma_z+d_0\sigma_0) \mathsf{U}$.
Without loss of generality, an experimentally accessible Hermitian operator acting on $n$ qubits can be written in a compact form as
\begin{eqnarray}
	O &=& \bigotimes_{i=1}^n(\mathbf{d} \cdot \boldsymbol{\sigma}+d_0\sigma_0)_{i} = \bigotimes_{i=1}^n[\mathsf{U}^{\dagger}(d\sigma_z+d_0\sigma_0) \mathsf{U}]_i, \notag\\
\end{eqnarray}
where the subscript $i$ labels the operators acting on the $i$th qubit, and the expectation of $O$ with respect to an $n$-qubit state $|\Psi\rangle$ can be measured as
\begin{eqnarray}
\left\langle \Psi \left| O \right| \Psi \right\rangle &=& \left\langle \Psi' \left| \bigotimes_{i=1}^n(d\sigma_z+d_0\sigma_0)_i\right| \Psi' \right\rangle\notag\\
&=&\prod_{i=1}^n \left( d\langle\sigma_z\rangle'+d_0\right)_i \notag\\
&\approx&\prod_{i=1}^n \left( d\frac{N'_0-N'_1}{N'}+d_0\right)_i, \label{eq: measurement}
\end{eqnarray}
where $|\Psi'\rangle = \bigotimes_{i=1}^n \mathsf{U}_i|\Psi\rangle$ and the prime symbol denotes the quantities with respect to $|\Psi'\rangle$.

For the readout state $\ket{\Psi_2}$ of Eq. \eqref{eq: Psi 2} in the main text, which consists of $n$ qubits in each of systems A and B, and one ancilla qubit labeled by $a$,  applying Eq. \eqref{eq: measurement} yields
\begin{eqnarray}
	&& \left\langle \Psi_2 \left| O \otimes O' \otimes \sigma_{x,y}\right|\Psi_2 \right\rangle \notag \\
	&\approx&\prod_{i=1}^n \left( d\frac{N'_0-N'_1}{N'}+d_0\right)_{i,A}  \left( d\frac{N'_0-N'_1}{N'}+d_0\right)_{i,B} \notag \\
	&&\times\left( d\frac{N'_0-N'_1}{N'}+d_0\right)_a, \label{eq: circuit_measure}
\end{eqnarray}
where $|\Psi'_2\rangle = \bigotimes_{i=1}^n \mathsf{U}_{i,A}\bigotimes_{i=1}^n \mathsf{U}_{i,B}\otimes\mathsf{U}_{a}|\Psi_2\rangle$.
Here, $n$ can take up to $\sim\!50$, because the scale of the current quantum platform can reach to about 100 qubits \cite{Google2023}.
Because of the direct-product form, the measurement for each qubit can be performed in parallel.

In the case of the nonreciprocal SSH model as an application in the main text, for the numerators on the left-hand side of Eq. \eqref{eqs: expectation}, we have $n=1$, $O=\boldsymbol{\sigma}$, and $O'=\sigma_0$, and Eq. \eqref{eq: circuit_measure} reduces to
\begin{eqnarray}
	&& \left\langle \Psi_2 \left| \boldsymbol{\sigma} \otimes \sigma_0 \otimes \sigma_{x,y}\right|\Psi_2 \right\rangle \notag \\
	&\approx&\left( \frac{N'_0-N'_1}{N'}\right)_{i,A}\left( \frac{N'_0-N'_1}{N'}\right)_a, \label{eq: circuit_measure_ssh}
\end{eqnarray}
where  $|\Psi'_2\rangle = \mathsf{U}_{A}\otimes\mathsf{U}_{a}|\Psi_2\rangle$ with $\mathsf{U}_{A,a}$ depending on the implementing operators, i.e., $\mathsf{U}_{A,a}(\theta)=\mathsf{Y}(-\pi/2)$ for $\sigma_x$, $\mathsf{X}(\pi/2)$ for $\sigma_y$, and $\mathsf{I}$ (no rotation) for $\sigma_z$.

For the denominators on the left-hand side of Eq. \eqref{eqs: expectation}, we have $n=1$ and $O=O'=\sigma_0$, and Eq. \eqref{eq: circuit_measure_ssh} reduces to
\begin{eqnarray}
	&& \left\langle \Psi_2 \left| \sigma_0 \otimes \sigma_0 \otimes \sigma_{x,y}\right|\Psi_2 \right\rangle \approx\left( \frac{N'_0-N'_1}{N'}\right)_a,
\end{eqnarray}
where  $|\Psi'_2\rangle = \mathsf{U}_{a}|\Psi_2\rangle$ with $\mathsf{U}_{a}$ taking the same form as for the numerators.

In a word, after obtaining the final state $|\Psi_2\rangle$, we need to first rotate it to $|\Psi'_2\rangle$ in the $(2n+1)$-qubit space according to the expected measurement operators; then, we count the number of times that $|0\rangle$ and $|1\rangle$ are detected for each qubit; and finally, we can obtain the target generalized expectation through Eq. \eqref{eqs: expectation} in the main text.

\section{Dilation method} \label{sec: dilation method}

\begin{figure}[tb]
    \includegraphics[width=\columnwidth]{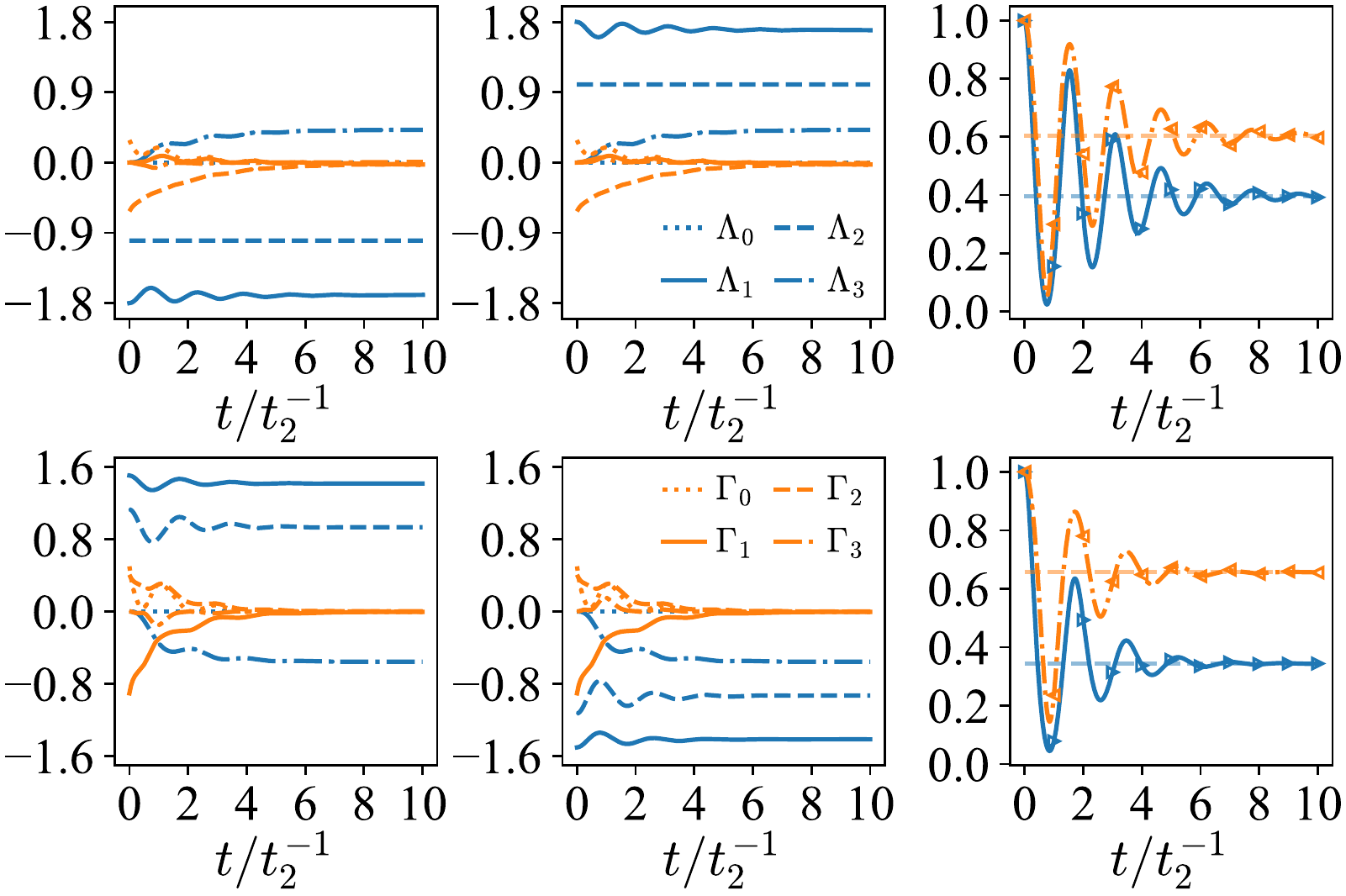}
	\caption{Comparison of the dilated method governed by $\mathcal{H}(t)$ and the nonunitary dynamics governed by $H(t)$.
		The upper and lower panels correspond to the circled point III in Fig. \ref{fig: winding number}(d) and the circled point II in Fig. \ref{fig: nb winding numbers}(d) of the main text with parameters $\{\eta_0,b\}=\{0.8,0.23\}$ and $\{0.7,0.35\}$, respectively.
		The left (middle) panels show the components of the dilated Hamiltonians $\mathcal{H}^{R}(t)$ [$\mathcal{H}^{L}(t)$].
        The legends in the middle panels also apply to the corresponding left panels.
		The right panels compare the dilation method (triangles) with the effective non-Hermitian one (lines) through the evolution of the population $|\langle0|\psi^{R,L}(t)\rangle|^2$.}
    \label{fig: dilation method example}
\end{figure}

In this appendix, we briefly outline the idea of the dilation method developed by Wu \textit{et al.} \cite{YWu2019} through two instances of the nonreciprocal SSH model under PBCs and OBCs.
This method can be implemented in several quantum systems, such as nitrogen vacancy (NV) centers \cite{YWu2019,WQLiu2020} and superconducting qubit systems \cite{Dogra2021}.

To find a state $\ket{\psi(t)}$ of a time-dependent non-Hermitian Hamiltonian $H(t)$, i.e.,
\begin{equation} \label{eq: Hs evo eqn}
    i \frac{d}{dt} \ket{\psi(t)} = H(t) \ket{\psi(t)},
\end{equation}
we can integrate it into a dilated state $\ket{\Psi(t)}$ of a composite system by introducing an ancilla qubit as follows:
\begin{equation} \label{eq: Psi}
    \ket{\Psi(t)} = \ket{\psi(t)} \otimes \ket{-} + e^{bt} \eta(t) \ket{\psi(t)} \otimes \ket{+},
\end{equation}
where $\eta(t)$ is an appropriate linear operator, $b$ is a coefficient to offset the amplitude of $|\psi(t)\rangle$ \cite{WQLiu2020}, and $\ket{-} = (\ket{0} - i\ket{1})/\sqrt{2}$ and $\ket{+} = (-i\ket{0} + \ket{1})/\sqrt{2}$ form an orthonormal basis of the ancilla qubit.
If we can attain the dilated state $\ket{\Psi(t)}$ in experiment, the state for which we aim, $\ket{\psi(t)}$, can be easily measured by postselecting the ancilla state $|-\rangle$.

The dilated state $\ket{\Psi(t)}$ can be evolved by a Hermitian time-dependent Hamiltonian $\mathcal{H}(t)$ in the composite system, i.e.,
\begin{equation} \label{eq: Hsa evo eqn}
    i \frac{d}{dt} \ket{\Psi(t)} = \mathcal{H}(t) \ket{\Psi(t)}.
\end{equation}
To make $\mathcal{H}(t)$ accessible in experiment (in the case of NV centers, for example), we can choose
\begin{equation} \label{eq: Hsa simplify}
    \mathcal{H}(t) = \Lambda(t) \otimes \sigma_0 + \Gamma(t) \otimes \sigma_z,
\end{equation}
where
\begin{eqnarray}
    \Lambda(t) &=& \left\{H'(t) + \left[i \frac{d}{dt} \eta(t) + \eta(t)H'(t)\right] \eta(t)\right\} M^{-1}(t), \notag\\
    \Gamma(t) &=& i\left[H'(t) \eta(t) - \eta(t) H'(t) - i \frac{d}{dt} \eta(t)\right] M^{-1}(t), \notag\\
\end{eqnarray}
with
\begin{eqnarray}
    H'(t)&=&H(t)-ibI, \quad \eta(t)=\sqrt{M(t)-I},\notag\\
    M(t) &=& \mathcal{T}e^{-i \int_0^t H'^\dagger(t) dt} M(0) \bar{\mathcal{T}} e^{i \int_0^t H'(t) dt}.
\end{eqnarray}
We choose $M(0)=\eta(0)^2+I$, where $I$ is the identity operator. The setting of $\eta(0)=\eta_0 I$ ensures $\det[M(t)-I]>0$ during the whole evolution.
Using $H'(t)=H(t)-ibI$ to generate $\mathcal{H}(t)$ can reduce the experimental difficulty \cite{WQLiu2020}, where $H(t)$ is just our target non-Hermitian Hamiltonian.

To demonstrate the validity of the dilation method in our examples in the main text, we generate the dilated Hamiltonian with $t_1/t_2=1.8$ ($1.6$) under PBCs (OBCs), $\delta/t_2=1/2$, and $k=\pi/2$ as the test bed.
    Choosing $\eta_0=0.8$ and $b=0.23$ under PBCs ($\eta_0=0.7$, $b=0.35$, and $\alpha=e^{i\pi/16}$ under OBCs), each component of $\mathcal{H}^{R,L}(t)$, defined as $\Lambda(t)/t_2 = \sum_i \Lambda_i(t) \sigma_i$, $\Gamma(t)/t_2 = \sum_i \Gamma_i(t) \sigma_i$, is plotted in the left and middle panels of Fig. \ref{fig: dilation method example}.
    To show the validity of the dilation method, the right panels of Fig. \ref{fig: dilation method example} compare the population evolutions $|\langle0|\psi^{R,L}(t)\rangle|^2$ generated by the dilation method of $\mathcal{H}^{R,L}(t)$ (triangles) and by the effective non-Hermitian method of $H(t)$ (lines), respectively, which shows that the dilation method is reliable.

\begin{acknowledgments}
We thank Dan-Bo Zhang, Yong-Li Wen, and Yu-Guo Liu for helpful discussions.
This work was supported by the National Natural Science Foundation of China (Grant No. 12074180), the Key-Area Research and Development Program of Guangdong Province (Grant No.~2019B030330001), the National Key Research and Development Program of China (Grant No.~2022YFA1405304), and the Guangdong Provincial Key Laboratory (Grant No.~2020B1212060066).
\end{acknowledgments}

% \nocite{*}

%

\end{document}